\documentclass[pdflatex,sn-nature]{sn-jnl}
\usepackage{graphicx}%
\usepackage{multirow}%
\usepackage{amsmath,amssymb,amsfonts}%
\usepackage{enumitem}
\usepackage{lmodern}
\usepackage{anyfontsize}
\usepackage{multirow}
\usepackage{amsthm}%
\usepackage{mathrsfs}%
\usepackage[title]{appendix}%
\usepackage{xcolor}%
\usepackage{textcomp}%
\usepackage[version=4]{mhchem}
\usepackage{chemformula}

\begin{document}
\author{Maryam M. Ahmed}
\affil{Physics Department, Faculty of Science, Zagazig University}
\supervision{Dr. Nagwa Abu-Elsaad}

\abstract{Ferrites, magnetic materials primarily composed of iron oxides, exhibit diverse magnetic behaviors, including diamagnetism, paramagnetism, ferromagnetism, antiferromagnetism, ferrimagnetism, and superparamagnetism. This paper explores the fundamental principles governing these magnetic phenomena, with a focus on the relationship between magnetic field strength ($H$), magnetic induction ($B$), and magnetic susceptibility ($\chi$). The study delves into the structural and electronic origins of magnetism in ferrites, highlighting the contributions of electron spin and orbital motions. The technological significance of ferrites, particularly in high-frequency applications, is examined through their classification based on crystal structures, including spinel, garnet, hexagonal, and orthoferrites. Special attention is given to the manufacturing processes and applications of ferrites in modern technology, such as their use in magnetic cores, sensors, and memory devices. The paper concludes with a discussion on the future directions in ferrite research and potential innovations in their applications.

}

\title[Exploring the Magnetic Behavior of Ferrites: From Diamagnetism to Superparamagnetism]{Exploring the Magnetic Behavior of Ferrites: From Diamagnetism to Superparamagnetism}

\maketitle

\section{Introduction}\label{sec1}
As early as 800 BC, magnetism was discovered in load stone, a naturally occurring substance that had been used for navigation. According to contemporary theory, magnetism exists within all materials, including semiconductors, insulators, and metals, though in a variety of ways. Additionally, the magnetizing properties of solids are essential, and investigation into them has provided profound insights into the underlying structure of many solids, both metallic and non-metallic. The magnetic field strength, also known as the externally applied field, is denoted by  H. The magnetic induction or magnetic flux density denoted by B reflects the amount of internal field strength inside a material as a consequence of the H field. Tesla units are used for B. The relationship between flux density and magnetic field strength is given by  

\begin{equation}
    \bf{B}= \mu\bf{H}
\end{equation} 
where $\mu$ indicates the permeability. The dimensions of permeability are measured in Henries per meter (H/m) or Weber per ampere meter $(Wb/Am)$. In vacuum 
\begin{equation}
    \bf{B} =\mu_0 \bf{H}
\end{equation}
where $\mu_0$ is the permeability of vacuum and it has a value $4 \pi  \times 10^{-7}(1.257\times10^{-6})\; H/m$
When a solid is subjected to a magnetic field and generated magnetic moment, it becomes magnetized. The magnetic dipole moment per unit volume is the magnetization vector M. Within the medium, there is magnetic induction as :
\begin{equation}
    \bf{B} = \mu_0\bf{H} + \mu_0 \bf{M}
\end{equation}
where $\mu_0\bf{M}$ is the result of medium magnetization and $\mu_0\bf{H}$ is caused by an external field.
Since the field induces magnetization, we can say that $\bf{M}$ is proportional to $ \bf{H}$, hence,
\begin{equation}
   \bf{M} = \chi \bf{H}
\end{equation}
The magnetic susceptibility of the medium is denoted by $\chi$ the proportionality constant $\chi$ which is defined as the amount of magnetization produced per unit applied magnetic field,i.e.$\chi=\bf{M}[Am^{-1}]/\bf{H}[Am^{-1}]$[unitless], is a parameter used to measure the state of magnetic materials.

In this case we have assumed $\chi$ to be isotropic. Nevertheless actual crystals are anisotropic therefore $\chi$ is a tensor. If Equation (4) substituted in Equation (3), we get:
\begin{equation}
    \bf{B} = \mu_0\bf{H} + \mu_0 \chi \bf{H} = \mu_0(1 + \chi)\bf{H}=\mu\bf{H}
\end{equation}
Where 
$$\bf{\mu}= \mu_0(1 + \chi)$$
and
$$ \mu_r = 1 + \chi$$

Based on modern theories, electron orbital and spin motions as well as nuclei's spins are responsible for the creation of magnetism in solids. The magnetic effects are created by electron motion, which is analogous to an electric current. Permanent electronic magnetic moments are created by the spin of unpaired valence electrons, which makes a significant contribution. A net non-zero magnetic moment can be produced by many of these magnetic moments aligning themselves in various directions. Consequently the nature of magnetization generated depends on the number of unpaired valence electrons present in the atoms of the solid and on the relative orientations of surrounding magnetic moments. The following six categories describe the magnetism inherent in solids:
\vspace{10 pt}
\begin{enumerate}[label = \roman*.]
\item Diamagnetism
\item  Paramagnetism
\item Ferromagnetism
\item Antiferromagnetism
\item Ferrimagnetismit
\item superparamagnetism
\end{enumerate}
\section{DIA, PARA, FERRI, ANTIFERRO, FERRO AND SUPERPARAMAGNETIC MATERIALS}

\subsection{Diamagnetism}

Diamagnetism is very lacking form of magnetism and is non-permanent and persists only in the event an external magnetic field is applied. It arises from an alteration in the electron's orbital motion when an external magnetic field is applied. The induced moment has a very little magnitude. The applied field's direction and the induced magnetic moment's direction are opposing. The relative permeability is somewhat smaller than unity. The magnitude of $B$ within the diamagnetic material is smaller than that in a vacuum, demonstrating a negative magnetic susceptibility. In cases where a diamagnetic material is set between the poles of a powerful electromagnet it is attracted towards regions at which field is weakened. Solid substances have a susceptibility of around $-10^{-5}$. All materials contain diamagnetism.
It can only be seen when all other forms of magnetism are completely gone due to its extreme weakness. It is noticed that ionic and covalent crystals exhibit diamagnetism. Other examples include metals like copper, gold, silver, and bismuth, organic solids like benzene and naphthalene, and atoms with rare gas configuration like argon, neon, and helium.

\begin{figure}[!htb]
    \centering
    \includegraphics[width=0.75\linewidth]{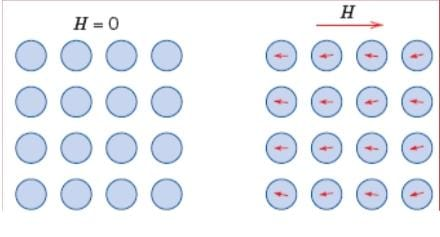}
    \caption{Atomic dipole configuration for a diamagnetic material without and with
magnetic field}
    
\end{figure}

\subsection{Paramagnetism}

When a substance experiences a permanent dipole moment due to partial cancellation of its electron spin and/or orbital magnetic moments, it is said to be paramagnetized. In the absence of a magnetic field, the orientation of these magnetic dipole moments is random, leading to no net magnetization. Once the magnetic field is applied, the dipole moments try to align with the magnetic field and the induced magnetization tends to increase the magnetic field . Because the alignment forces are smaller than the forces from thermal motion that attempt to disrupt the alignment, paramagnetism is often rather weak.
The paramagnetism is sensitive to the temperature, the lower the temperature the stronger the effect. It seems that there is greater alignment at low temperatures since the influence of thermal motion is reduced
Since ions and atoms with an odd number of electrons must be paramagnetic due to their angular momentum, paramagnetic states are seen in these materials.paramagnetic has positive and very little susceptibility on a range of $10^{-3}$to $10^{-5}$. This behavior in the presence of a very weak magnetic field is explained by Curie's law
$$\chi=M/H=C/T$$
Here $C$ is known as the Curie constant.
\begin{figure}[!htb]
    \centering
    \includegraphics[width=1\linewidth]{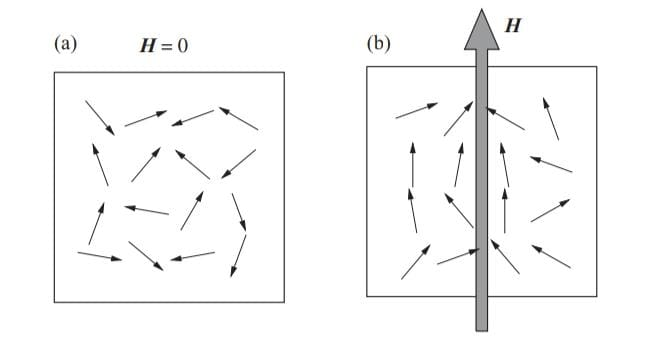}
    \caption{Schematic of the alignment of magnetic moments in a paramagnetic
material: (a) shows the disordered arrangement in the absence of an external field,
and (b) shows the response when a field of moderate strength is applied.}
    
\end{figure}

\subsection{Ferromagnetism}

In the absence of an external magnetic field, some metallic materials such as Fe, Co, and Ni possess a permanent magnetic moment and show extremely strong magnetizations.
 In ferromagnetic materials, atomic magnetic moments resulting from uncancelled electron spins as a consequence of electron structure give rise to permanent magnetic moments. Above a specifically temperature known as Curie temperature $(Tc)$ the ferromagnetic material demonstrate paramagetism
\begin{figure}[!htb]
    \centering
    \includegraphics[width=0.5\linewidth]{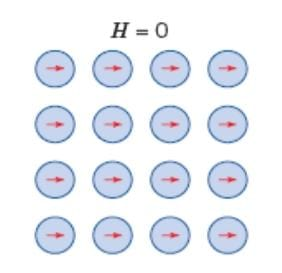}
    \caption{The orientation of atomic dipoles for a ferromagnetic material}
    
\end{figure}

Weiss' theory of ferromagnetism states that a sample of ferromagnetic material has several tiny, spontaneously magnetic areas termed domains. The vector sum of the magnetic moments of the individual domains determines the overall amount of spontaneous magnetization of the specimen. The exchange field, $B_E$, which tends to generate a parallel alignment of the atomic dipoles, is what causes the spontaneous magnetization of each domain. It is assumed that the field $B_E$ is proportionate to each domain's magnetization M, thus that 
$$B_E=\lambda{M}$$
where $\lambda$ is a temperature-independent constant known as the Weiss-field constant.
In general, the applied field B is not as strong as the field $B_E$, which is also referred to as the molecular field. For iron, $B_E$ equals 1000 Tesla. thereby, the effective magnetic field on atom transforms into
$$B_{eff}=B+B_E=B+\lambda M$$
When an external magnetic field is present, the ferromagnetic material becomes magnetized. This is caused by two factors:
\begin{enumerate}
    \item The size of the domains with favorable orientations grows, while the size of the domains with unfavorable orientations decreases in response to the applied field.
    \item The rotation of several domains' magnetization orientations along the applied field's direction.
\end{enumerate}

\begin{figure}[!htb]
    \includegraphics[width=1\linewidth]{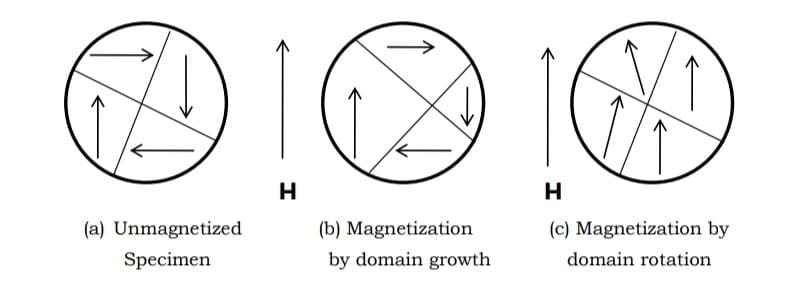}
    \caption{Two fundamental processes of magnetization in ferromagnetic materials}
    \label{ferro}
    
\end{figure}

\noindent These two magnetization processes are illustrated in figure. \ref{ferro}

Typically, in weak fields, domain boundary displacement causes the magnetization, increasing the size of the favorably oriented domains. In strong field the magnetization takes place by the rotation of domains. When the field is removed the domain boundary do not move totally back to their initial position, and the material stays still magnetized. At this point the material seems permanent magnet. The ferromagnetic material becomes paramagnetic with individual atomic dipoles when the domains split apart at high temperatures.
The susceptibility of a ferromagnetic material follows the Curie-Weiss equation above the Curie temperature, $T_C$. 
$$\chi=\frac{C}{T-T_c}$$

Here C is the Curie- Weiss constant

\subsection{Antiferromagnetism:}

Beyond ferromagnetic materials, other materials also exhibit this phenomenon of magnetic moment coupling between neighboring atoms or ions. The name "antiferromagnetism" refers to a group in which this coupling produces an antiparallel alignment, or the alignment of nearby atoms' or ions' spin moments in completely opposite directions.Manganese oxide [MnO] crystals were the first materials that showed this type of magnetism. The material does not show any magnetization when there is no external magnetic field because the magnetic moments of the adjacent particles cancel each other out. Nevertheless, when a field is applied, little magnetization becomes apparent in the direction of the field and gets stronger further with temperature .
Below what is known as the Neel temperature $T_N$, antiferromagnetism takes place.Thermal energy is enough above the Neel temperature  $T_N$  to cause the atomic moments oriented in the opposite direction to become randomly distributed, which eliminates their long-range order. In this situation, the material exhibits paramagnetic behavior.
\begin{figure}[!htb]
    \centering
    \includegraphics[width=0.5\linewidth]{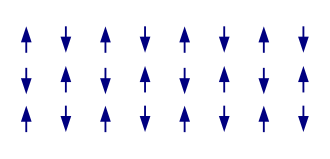}
    \caption{The arrangement of the atomic magnets in an antiferromagnetic material}
    
\end{figure}

\subsection{Ferrimagnetism}

\begin{figure}[!htb]
    \centering
    \includegraphics[width=1\linewidth]{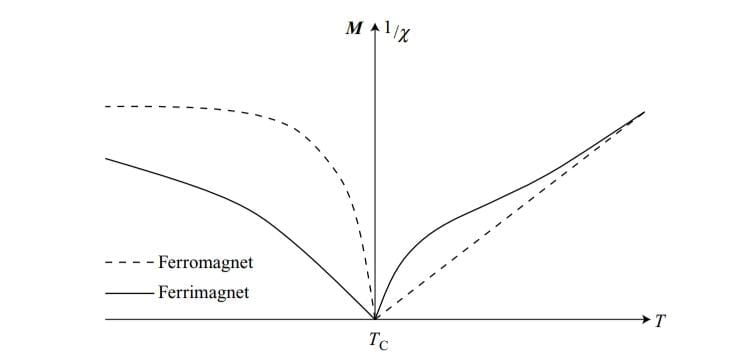}
    \caption{Comparison of magnetization and inverse susceptibility in typical ferri and ferromagnets.}
    \label{ferri}
\end{figure}

Similar to ferromagnets, ferrimagnets display spontaneous magnetization even in the absence of an applied field below a certain temperature, Tc. Nonetheless, a typical ferrimagnetic magnetization curve differs greatly from a ferromagnetic curve in shape, as seen in Fig. \ref{ferri}. In contrast to most ferromagnets, which are metals, ferrimagnets are electrically insulators simply because they are ionic solids. With the exception of the two sublattices' varying magnetization magnitudes, which provide a non-zero net magnetization value, ferrimagnetism and antiferromagnetism are actually the same 

Materials like ferrites, which are fundamentally the oxides of several metal elements, exhibit this type of magnetism and these materials may be described using the chemical formula $MFe_2O_4$, since M can be any one of different metallic elements .The most prevalent example is that of ferrous ferrite, frequently referred to as magnetite $Fe_3O_4$ or $FeO$.$Fe_2O_3$.
\begin{figure}[!htb]
    \centering
    \includegraphics[width=0.5\linewidth]{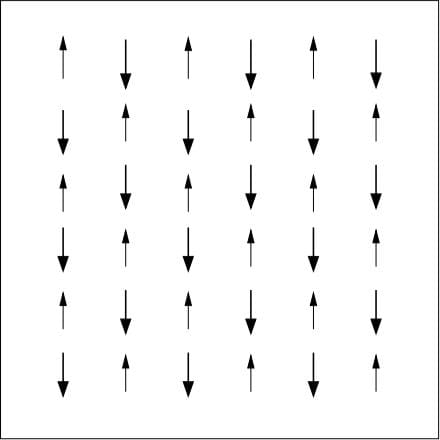}
    \caption{Ordering of magnetic ions in a ferrimagnetic lattice}
    
\end{figure}

\subsection{\large{Ferrites}}

The materials referred to as ferrites are the most significant ferrimagnets in terms of technology. Ferrites are electrically insulators and ferrimagnetic transition-metal oxides. They therefore find use in circumstances where the electrical conductivity exhibited by the majority of ferromagnetic materials would be harmful. Because an ac field does not create undesired eddy currents in an insulating material, they are frequently utilized in high-frequency applications Ceramic processing procedures are mostly used in the manufacturing of ferrites. For instance, powdered NiO and $Fe_2O_3$ are combined, shaped, and heated to create $NiO.Fe_2O_3$. One benefit of this technology is that it's simple to regulate the magnet's form by selecting the appropriate mold.
There are two primary families of magnetic ferrites, each having a distinct crystal structure:

1. Cubic. The usual formula for these is MO.$Fe_2O_3$, where M is an ion of a divalent metal, such as Mn, Ni, Fe, Co, or Mg. While the other cubic ferrites are magnetically soft, cobalt ferrite, $CoO.Fe_2O_3$, is magnetically hard. These ferrites are both ancient and modern magnetic materials as magnetite $Fe_3O_4$ , also known as iron ferrite, is the earliest magnetic substance ever discovered by humans and is referred to as the "lodestone" of antiquity.

2. Hexagonal. The most significant members of this category are the magnetically hard barium and strontium ferrites, BaO.6$Fe_2O_3$ and SrO.6$Fe_2O_3$.

\subsection{superparamagnetism}
Small ferromagnetic or ferrimagnetic nanoparticles exhibit a form of magnetism known as superparamagnetism (SPM). This suggests sizes that, depending on the substance, range from a few nanometers to a few tenths of a nanometer. These nanoparticles are single-domain particles as well. The overall magnetic moment of the nanoparticle may be roughly expressed as one large magnetic moment made up of all the individual magnetic moments of the constituent atoms.
Nanoparticles frequently exhibit a preference for the direction that their magnetization aligns with. It is claimed that these nanoparticles exhibit an anisotropy in these directions. Uniaxial anisotropy is used to describe it when there is just one primary direction.
Uniaxially anisotropic nanoparticles randomly reverse the direction of their magnetization. Thermal energy is what causes this phenomenon. The relaxation time indicates the typical amount of time needed to complete such a reverse:
$$\tau = {\tau_0}exp\left(\frac{\Delta E}{KT}\right)$$
where :
$\tau$is the investigated material's characteristic length of time. 
$\Delta{E}$ is the energy barrier the magnetization flip has to overcome by
thermal energy. K is the Boltzmann constant and T is the temperature
However, the energy barrier $\Delta{E}$ and temperature T are not the only factors that affect the detection of nanoparticles in a superparamagnetic state: Every experimental method has a unique measurement time $\tau_m$. Two possible outcomes can arise based on measurement time :

$\tau_m<<\tau$ : The measurement time is significantly smaller than the average time -between flips. This places the particles in a clearly defined state that is commonly known as the system's blocked state.

$\tau_m>>\tau$ : On the other hand, the measurement time may be significantly longer than the average time between flips. This suggests that a fluctuating condition with many unresolved magnetization spin orientations is really observed by the experiment. In the absence of any external field, a time-averaged net moment of zero is measured. A system is said to be in its superparamagnetic condition in this circumstance. 
The temperature that separates the blocked state from the superparamagnetic state is known as the blocking temperature, or $T_B$. This indicates that at the blocking temperature, $\tau_m=\tau$
$$T_B=\frac{\Delta E}{KLn(\frac{\tau_m}{\tau_0})}$$
As a result, the following characteristics allow us to differentiate the two states:

$\tau_m<<\tau$ or $T_B>T$ is the blocked state.

$\tau_M>>\tau$ or $T_B<T$ is the superparamagnetic state.

\begin{figure}[!htb]
    \centering
    \includegraphics[width=0.5\linewidth]{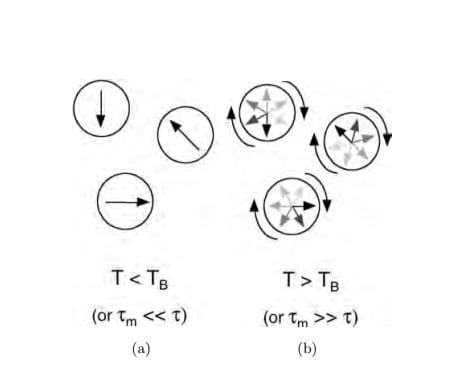}
    \caption{Case (a): The relaxation time is significantly longer than the measurement time, One observes a clearly defined state. Case (b): The relaxing time is significantly smaller than the measurement time .The superparamagnetic condition,characterized by a time-averaged net moment of zero, is caused by the magnetization's fluctuating state.  }
\end{figure}

What consequences arise from such superparamagnetic states? The net moment is 0 in the absence of an external magnetic field. The nanoparticles behave similarly to a paramagnet when an external field is applied, which is why the word "paramagnetism" is used. The only difference is that the nanoparticles' magnetic susceptibility is significantly greater, which is why the name "super" is used.
To be clear: Any ferromagnetic or ferrimagnetic substance may often exhibit paramagnetically. This is starting at what is known as the Curie temperature, or $T_C$, and going higher. Nevertheless when superparamagnetic behavior is seen below the Cure temperature, an other explanation is required.

\section{FERRITES}

Fundamentally, ferrites are extremely hard and brittle ceramic materials with a dark grey or black appearance. Ferrites are magnetic materials made of oxides with ferric ions as the primary constituent component (the term "ferrite" is derived from the Latin "ferrum," which means iron) and are categorized as magnetic materials due to their ferrimagnetic properties .Ferrites have exceptional magnetic and electrical properties. Among their many advantages are their wide frequency range, form flexibility, affordability, high electrical resistivity, minimal eddy current loss, high permeability, and time-temperature stability. The high-temperature solid-state reaction method, sol-gel method, coprecipitation, pulsed laser deposition, high-energy ball milling, and hydrothermal process can all be used to create ferrites in powder or thin film form. 
In order to create a ferrite core, a combination of powders comprising the component raw materials is pressed into the desired form before being sintered into a ceramic component. The interactions between metallic ions that occupy certain places in relation to the oxygen ions in the oxide's crystal structure give birth to the magnetic characteristics.
\subsection{Classification of Ferrites}
\begin{figure}[!htb]
    \centering
    \includegraphics[width=1\linewidth]{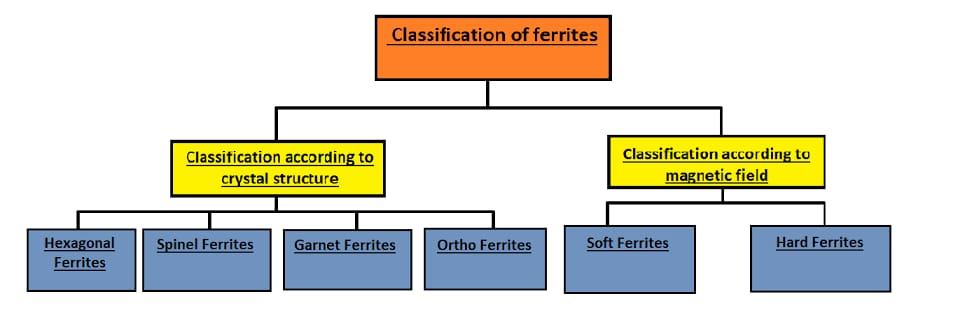}
    \caption{Classification of ferrites}
    \label{class}
    \end{figure}

Ferrites are categorized in the following ways based on their magnetic characteristics and crystal structures: Fig. \ref{class}

\subsection{Based on Crystal Structure}
The crystal structures of ferrites vary from one another. Ferrites may be divided into four categories according to their crystal structures: spinel, garnet,hexagonal, and Orthoferrites ferrite.

\subsubsection{Spinel Ferrites [Cubic Ferrites]}

\begin{figure}[!htb]
        \centering
        \includegraphics[width=1\linewidth]{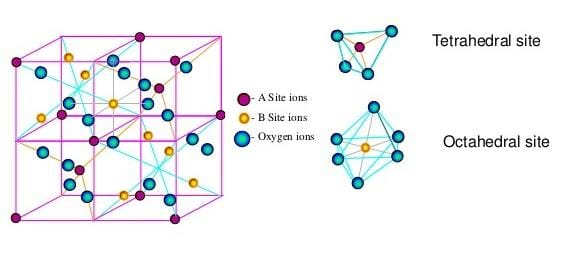}
        \caption{Crystal structure of a spinel ferrites}
        \label{spinel}
    \end{figure}

A spinel is a cubic structure with the theoretical equation A$B_2O_4$, where A denotes a divalent cation and B denotes a trivalent cation.On the other hand, the formula $MFe_2O_4$, where M denotes divalent metal ions such as Fe, Cu, Ni, Mg, Mn, Co, Zn, Cd, etc., describes Spinel ferrites.Al, Cr, Ga, In, and other trivalent ions can be used in place of the $Fe^{3+}$.The word "ferrites" was applied to these minerals because of their structural similarity to the naturally occurring mineral spinel, $MgAl_2O_4$.
The most significant class of magnetic materials, spinel ferrites, have several intriguing uses.
Due to their exceptional magnetic characteristics, spinel ferrites, which are magnetically soft and a better option than metallic magnets like Fe and layered Fe-Si alloys, perform better . They can also offer minimal magnetic losses as well as excellent electrical resistance. The two most common types of ceramic magnets in the spinel ferrite family are manganese-zinc ferrites and nickel-zinc ferrites.
Their high magnetic permeability, high electrical resistivity, and potential for wide-spectrum alteration of inherent characteristics make them attractive ceramic materials. 
Spinel ferrite crystallizes in the cubic structure and is formed of a close-packed oxygen anions organization in which 32 oxygen ions form the unit cell.
Two types of spaces are left between the anions as a result of the packing of these anions in a face-centered cubic (FCC) arrangement: tetrahedrally coordinated sites (A), which are surrounded by the four nearest oxygen atoms, and octahedrally coordinated sites (B), which are surrounded by the six nearest neighbor oxygen atoms.with two interstitial sites, tetrahedral (A) and octahedral (B), the crystal structure is cubic with spinel-type, or $MgAl_2O_4$. Since the unit cell consists of eight units (cubes), it may be expressed as $M_8Fe_{16}O_{32}$,which was initially created by Nishikawa and Bragg
The spinel ferrite structure is
shown in Fig. \ref{spinel}

(a) Classification:

Spinel ferrites are divided into three forms according to the distribution of cations in the two main sites, tetrahedral site (A) and octahedral site (B).

\begin{figure}[!htb]
    \centering
    \includegraphics[width=0.75\linewidth]{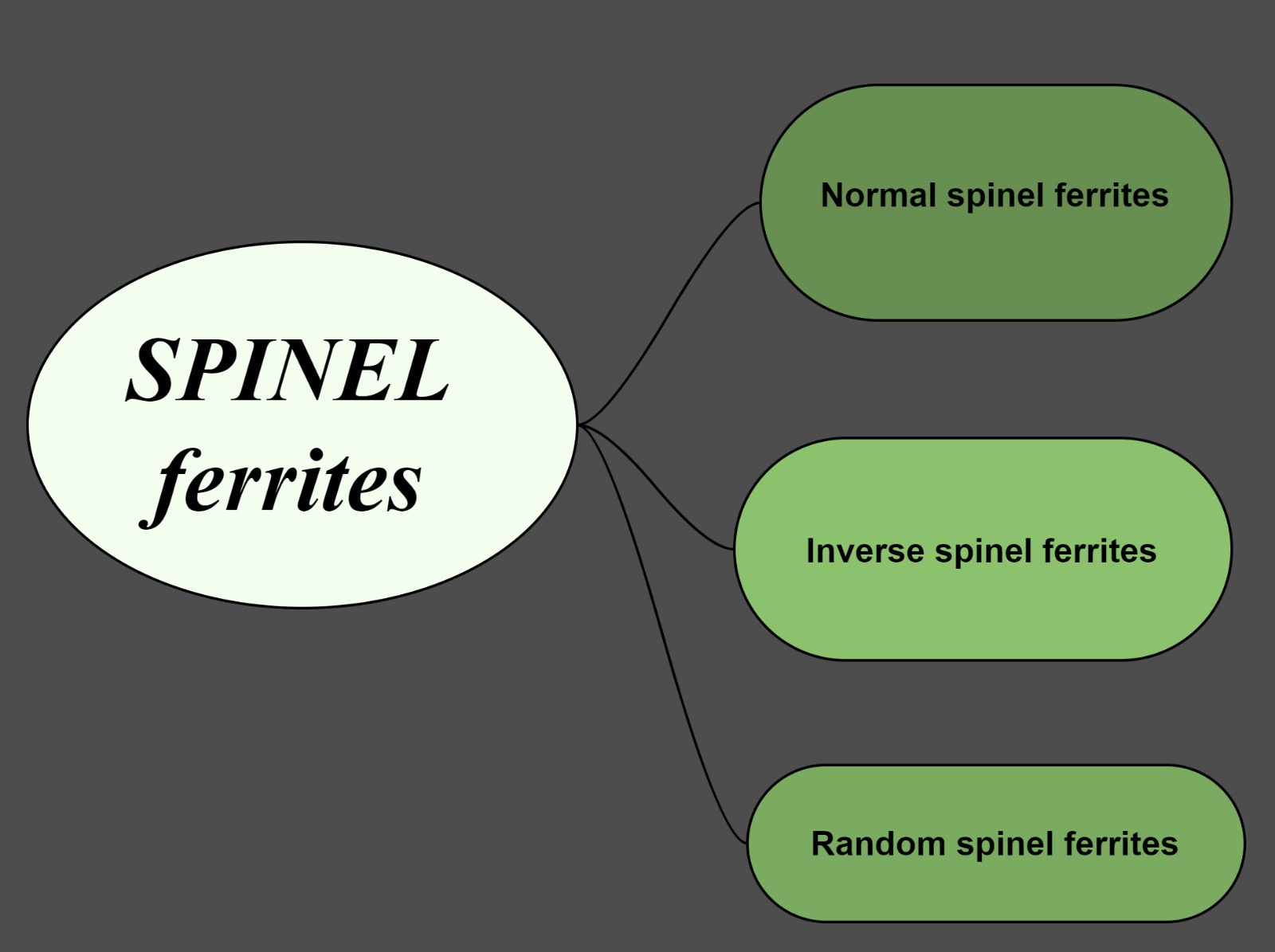}
    \caption{Types of spinel ferrites}
\end{figure}

(i) Normal spinel ferrites:

When $2Fe^{3+}$ ions are at the octahedral (B) site and divalent metal ions are at the tetrahedral (A) site, the ferrite is referred to as normal spinel. Commonly, cubic close-packed oxides with four octahedral and eight tetrahedral sites per formula unit make up normal spinel structures. Zinc (Zn$Fe_2O_4$) and cadmium (Cd$Fe_2O_4$) ferrites, where the divalent metallic ions $Zn^{2+}$ or $Cd^{2+}$are at the (A) site and $Fe^{3+}$ ions are at the (B) site, are the greatest examples of typical spinel ferrites. In general, $$(M)_A [Fe_2]_BO_4$$ can be employed to represent the cation distribution.

(ii) Inverse spinel ferrites:

One trivalent ferric ion, $Fe^{3+}$, is located at the tetrahedral (A) site in inverse spinel ferrite, whilst the remaining trivalent ferric ions, $Fe^{3+}$, and the divalent metallic ions, $M^{2+}$, are located at the (B) site. This group really includes the majority of simple ferrites, such as cobalt ferrite (Co$Fe_2O_4$), nickel ferrite (Ni$Fe_2O_4$), and ferric oxide ($Fe_3O_4$) . In inverse spinel ferrite, the cation distribution is represented as $$ (Fe)_A [M Fe]_BO_4$$.

(iii) Random spinel ferrites:

The ferrite is known as random spinel ferrite when the divalent metal ions $M^{2+}$ and trivalent $Fe^{3+}$ ions are dispersed at both the tetrahedral (A) and octahedral (B) sites. Additionally, of the random spinel ferrites, copper ferrite ($CuFe_2O_4$) is the most well-known example.Ionic radii, their electronic configuration, and the electrostatic energy of the lattice all play a part in the delicate balance of contributions that determine the distribution of ions between two types of sites . Random spinel's cation distribution is often represented by $$(M_{1-x}Fe_x)_A [M_xFe_{2-x}]_BO_4$$.

\subsubsection{Hexagonal Ferrites}
Went, Rathenau, Gorter, and Van Oostershout (1952) and Jonker, Wijn, and Braun (1956) made the initial identification of hexaferrite. Rhombohedral ferromagnetic oxides are another name for hexagonal ferrites. The formula $MFe_{12}O_{19}$, where M is an element such as barium (Ba), strontium (Sr), calcium (Ca), or lead (Pb), is often used to identify hexagonal ferrites.These ferrites have a strong coercive permanent (hard) magnet-i.e.with typical coercivities of around 200 kA/m- characteristic with oxygen ions having a closely packed hexagonal crystal structure. Because it is difficult to readily switch the axis of magnetization in hexagonal ferrites, they are known as hard ferrites.In addition, unlike cubic ferrites,hexagonal ferrites have several types, including M-type hexaferrite, W-type hexaferrite, Y-type hexaferrite, Z-type hexaferrite,and U-type hexaferrite,as shown in Table. \ref{hexa}.
\begin{table}[!htp]
\centering
\begin{tabular}{ |p{1.5cm}||p{2.5cm}|p{3.5cm}|  }
\hline
 Sr. no.& Hexaferrite &Chemical formula\\
 \hline
 1  & M-type    &$AFe_{12}O_{19}$\\
 \hline
 2&   Y-type & $A_2Me_2Fe_{12}O_{22}$\\
 \hline
 3 &W-type & $AMe_2Fe_{16}O_{27}$\\
 \hline
 4    &X-type& $A_2Me_2Fe_{28}O_{46}$\\
 \hline
 5&   U-type  & $A_4Me_2Fe_{36}O_{60}$\\
 \hline
 6& Z-type  & $A_3Me_2Fe_{24}O_{41}$ \\
 \hline
\end{tabular}
\caption{Different types of hexaferrites.}
\label{hexa}
\end{table}

A few of instances of hexaferrite include strontium ferrite ($SrFe_{12}O_{19}$) and barium ferrite ($BaFe_{12}O_{19}$). In terms of technology and commerce, hexagonal ferrite is now a very significant material.
Of the hexagonal ferrites, barium ferrite ($BaO.6Fe_2O_3$) is the most significant.
The crystal structure of barium ferrite is hexagonal magnetoplumbite (Fig.12).
The magnetoplumbite structure is made up of four building blocks, denoted S, $S^{*}$, R, and $R^{*}$ in the figure, and has ten oxygen layers in its primary unit cell. The S and$S^{*}$ blocks are spinels with two oxygen layers and six $Fe^{3+}$ ions. Two of the $Fe^{3+}$ ions are in tetrahedral sites, having opposite spin direction to the octahedral iron ions, while the remaining four are in octahedral sites with their spins aligned parallel to each other (say up-spin). 
\begin{figure}[!htb]
    \centering
    \includegraphics[width=1\linewidth]{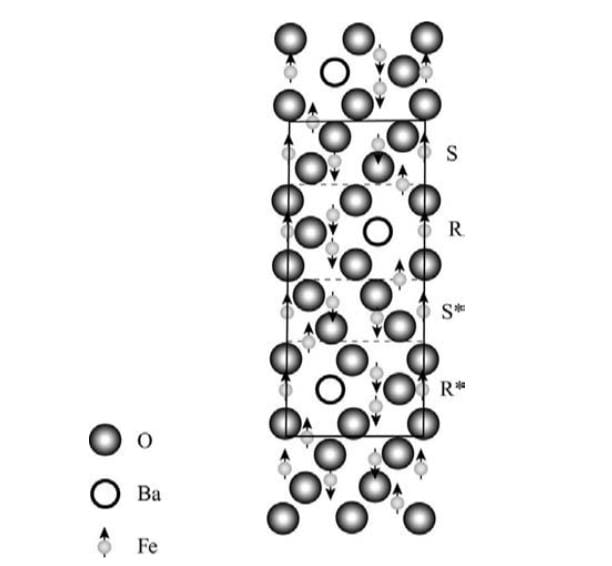}
    \caption{ Structure of barium ferrite.}
    
\end{figure}

The S and $S^{*}$ blocks are equivalent nevertheless rotated $180^{\circ}$ with respect to each other. The R and $R^{*}$ blocks are made up of three oxygen layers, with a barium ion taking the place of one of the oxygen anions in the middle layer. Each R block consists of six $Fe^{3+}$ ions, five of which are in octahedral sites with three up-spin and two down-spin, and one of which is coordinated by five $O^{2-}$ anions and has up-spin. Each unit cell has a net magnetic moment of 20$\mu$B.
Similar to cubic ferrites, they are inexpensive to make using ceramic processing techniques and are easily powdered and shaped into whatever shape one wants.

\subsubsection{Garnet Ferrites:}
Researchers' attention has recently been drawn to garnet-based nanoferrites due to their wide range of applications. In simple terms, garnets are minerals with the general formula $X_3Fe_5O_{12}$, which contains two magnetic ions: an iron ion and a rare earth element (where "X" is an element such as Sm, Eu, Gd, Tb, Dy, Er, Tm, Lu, and Y). Garnets have dodecahedral (12-coordinated) sites in addition to tetrahedral and octahedral sites which are magnetically hard Their unit cell has a cubic form with edges that are around 12.5 Å long. One well-known garnet is yttrium iron garnet, or $Y_3Fe_5O_{12}$.The garnets' crystal structure is orthorhombic, with oxygen polyhedral surrounding the cations. However, trivalent cations, such as Fe3+ and rare earth, occupy tetrahedral (d), octahedral (a), or dodecahedral—a 12-sided distorted polyhedral (c) site.
In particular, the net magnetic moment is antiparallel to the rare earth ions on the "c" sites, as well as the interaction between the tetrahedral and octahedral sites is antiparallel.
Among all crystal structures, garnet structure is one of the most complex, making it challenging to depict in two dimensions while include all 160 ions in the unit cell.
Fig.13 only displays an octant of a garnet structure, which just displays the cation locations, for the reason of simplicity. The garnet structure is consists of a combination of octahedral (trivalent cation surrounded by six oxygen ions), tetrahedral (trivalent cations surrounded by four oxygen ions), and 12-sided polyhedral dodecahedral (trivalent cations surrounded by 8 oxygen atoms) sites, the orientations of which are shown in Fig.12. The finding of a complicated crystal structure is noteworthy due to its applicability in memory devices.
\begin{figure}[!htb]
    \centering
    \includegraphics[width=0.8\linewidth]{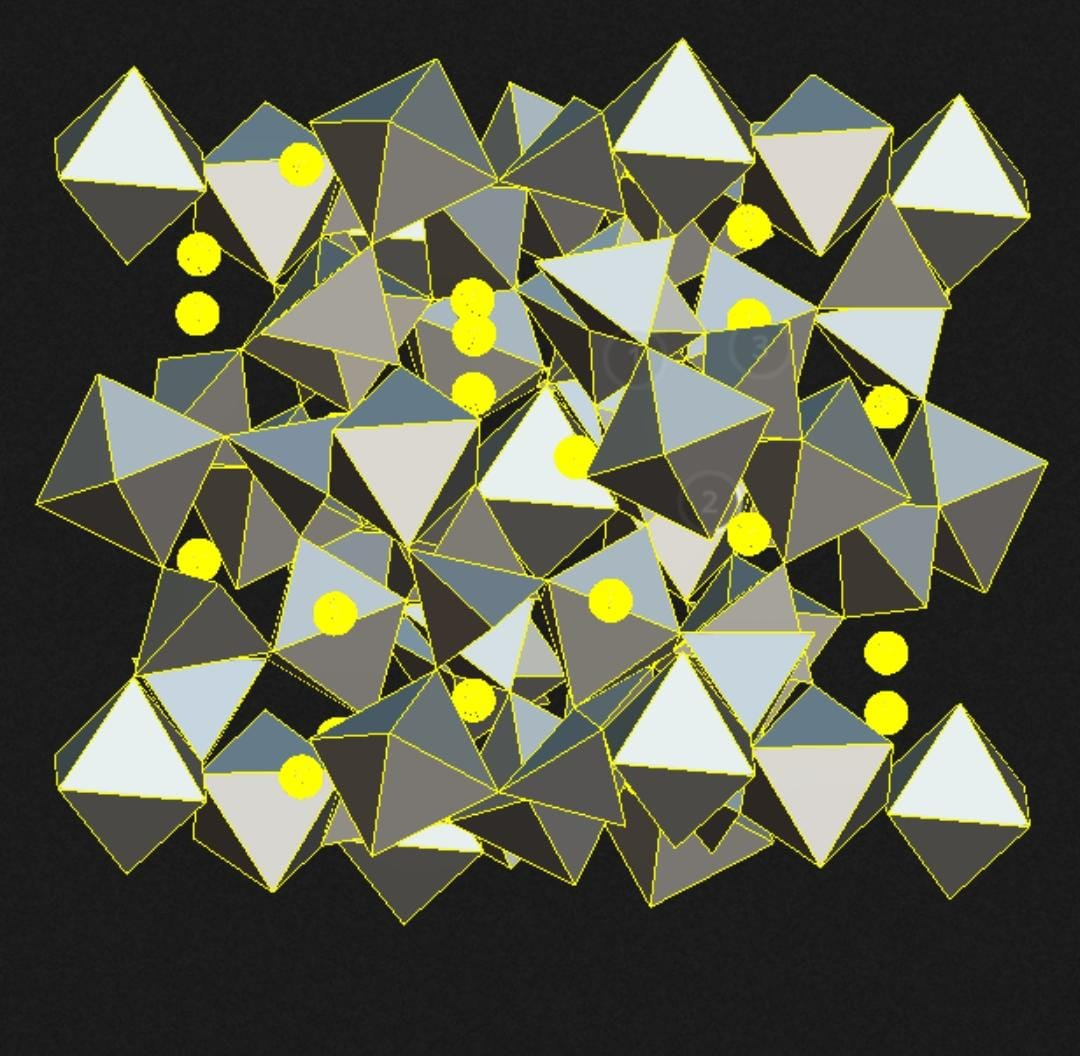}
    \caption{Crystal structure of Pyrope (Garnet) in three dimensions}
    
\end{figure}

 \begin{figure}[!htb]
    \centering
    \includegraphics[width=0.8\linewidth]{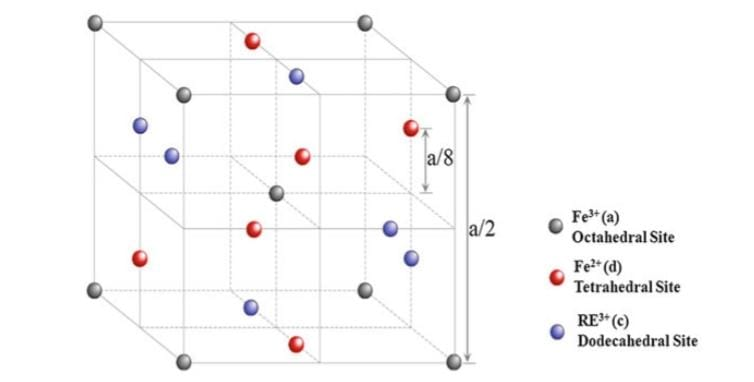}
    \caption{Schematic octant of a garnet crystal structure with only cation positions}

\end{figure}

\begin{figure}[!htb]
        \centering
        \includegraphics[width=0.8\linewidth]{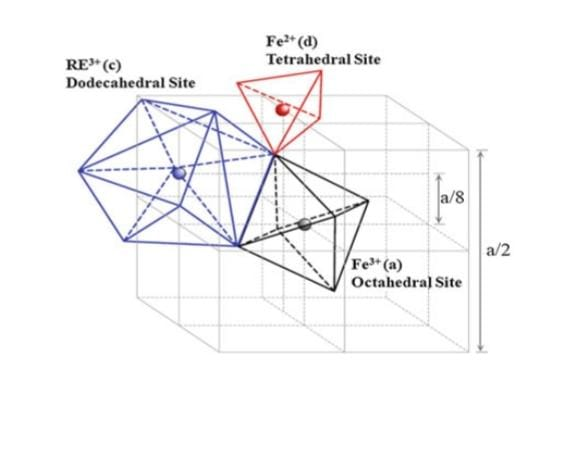}
        \caption{An octant of a
garnet crystal structure}

    \end{figure}

\subsubsection{Orthoferrites:}
Orthoferrites are denoted by the symbol MFeO3, in which M stands for one or more rare earth elements. Orthoferrites are lightly ferromagnetic and have an orthorhombic crystal structure. Examples of orthoferrite materials are dysprosium orthoferrite ($DyFeO_3$), lanthanum orthoferrite (LaFe$O_3$), and praseodymium orthoferrite (PrFe$O_3$). Orthoferrites have application in gas separators, catalysts, spin valves, magneto-optic materials, optical Internet, communication systems, cathodes in solid oxide fuel cells, and electrical circuits. In Fig. 17, the perovskite structure is illustrated. Tiny trivalent or tetravalent metal ions (B) occupy the cube's core, whereas large divalent or trivalent metal ions (A) occupy the cube's corners. On the cube's faces, the oxygen ions are positioned in the center
 \begin{figure}[!htb]
        \centering
        \includegraphics[width=0.8\linewidth]{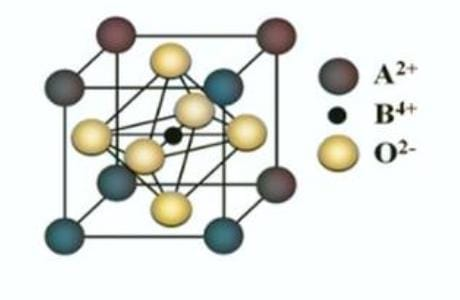}
        \caption{Structure of an orthoferrite perovskites}

    \end{figure}

\subsection{According to Magnetic Properties}
Ferrites demonstrating spontaneous magnetization are classified into two groups based on their magnetic properties: \textbf{soft ferrites and hard ferrites}.
\subsubsection{Soft Ferrites:}

Ferrimagnetic materials that lose their magnetism after being magnetized are known as soft ferrites. Soft ferrites have little energy losses during magnetization, a short and tiny hysteresis loop, and a sharply rising magnetization curve. These materials are easily magnetized and demagnetized and may be created by heating and slowly cooling. Soft ferrites are ceramic insulators that have the cubic crystal structure $MFe_2O_4$, where M stands for transition metal ions like Zn, Ni, and Mn. Low eddy current losses, low retentivity and coercivity, and high susceptibility and permeability (Nickel Zinc ferrite, for instance, has 1.26 ×$ 10^{-5}$-2.89 × $10^{-3}$ H/m) are characteristics of soft ferrites .Moreover, soft ferrites with cubic spinel structures known as NiZn, MnZn, while MgMnZn ferrites, and soft ferrites with garnet structures known as YIG. Soft ferrites are therefore utilized in electromagnets, computer data storage, transformer cores, telephone signal transmitters and receivers, and other devices. 
\subsubsection{hard ferrites:}
Because hard ferrites retain their magnetism even after being magnetized, they are also referred to as permanent magnetic materials. Hard ferrites are ferrimagnetic materials characterized by a large hysteresis loop, a gradually increasing magnetization curve, as well as large energy losses during the magnetization process. Hard ferrites are difficult to magnetize and demagnetize and can be created by heating and abruptly cooling. Iron and barium or strontium oxides combine to generate hard ferrites.Furthermore, the structure of hard ferrite is hexagonal. 
Hard ferrites have strong retentivity and coercivity, high eddy current losses, low susceptibility and permeability, and high saturation flux density. Hard ferrites are therefore utilized in loud speakers, DC magnets, permanent magnets, etc. 

\begin{figure}[!htb]
    \centering
    \includegraphics[width=1\linewidth]{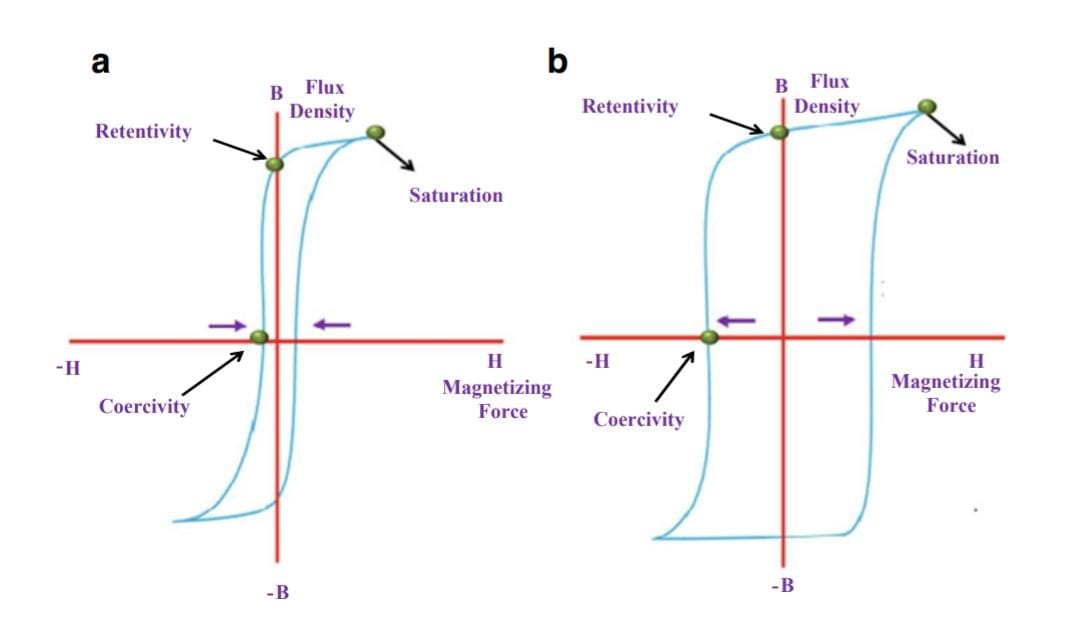}
    \caption{shows the hysteresis loop[M-H]of (a) soft ferrite,( b) hard ferrite}
    
\end{figure}
\subsubsection{Fundamental Applications Based on the Properties} 
As a result of their higher resistivity, reduced cost, ease of production, and superior magnetization properties, ferrites are thought to be more effectively magnetic materials than pure metals. Ferrites are widely employed in satellite communication, radar, bubble devices, computer memory cores, audio-video and digital recording, and microwave devices. Applications for ferrite are several, ranging from radio frequencies to microwaves.  It is utilized in radio reception antenna cores, TV image tube fly back transformers, broad band transformers, mechanical filters, ultrasonic generators, and isolators, moderators, and phase shifters. 
Ferrite is employed in computers, control devices, and telephone exchanges these days. 
 Additional applications include magnetrons, couplings, and zero friction bearings in aviation and space technology; speakers and motors for household appliances; sensors, routers, and converters in electrotechnology; and driving engines, igniters, ventilation systems, speed indicators, and existing power stations in electric drive trains. Clocks, ear buds, and magnetic locks are used in medical; in manufacturing, there are magnetic prostheses, cancer cell artificial organs, nuclear magnetic resonance apparatus, separators, and audiovisual analogue devices.

\section{Nature Of Domains}
The theory of magnetic domains inside solids was first presented by Weiss in 1907.Any ferromagnetic material, according to Weiss, is made up of tiny volume regions where all magnetic dipole moments are mutually aligned in the same direction. Every one of these areas is magnetized to its saturation magnetization and is typically referred to as a domain.The domain walls or boundaries that divide adjacent domains cause the magnetization to progressively change to a different direction. Bloch wall is the term for this wall or boundary. A portion of an unmagnetized domain contains several domains that are all spontaneously magnetized, but the orientation of the magnetization in each domain varies randomly. Each domain has around $10^{17}$–$10^{21}$ atoms. In the event of an external magnetic field, these domains align.
When an external magnetic field is present, the energy of the domains with net magnetic moments parallel to the field's direction decreases while the energy of the domains without such moments increases. If every domain is parallel to the applied magnetic field, the energy of the crystal can be reduced.
Here are the methods via which this can be acquired:

(i) A domain's magnetization direction can change all at once.

(ii) A favorably oriented domain expands at the expense of less favorably oriented domains. (Fig.18)

(iii) The magnetization rotates in the direction of the applied magnetic field under strong magnetic fields (Fig.18 )
\begin{figure}[ht]
    \centering
    \includegraphics[width=1\linewidth]{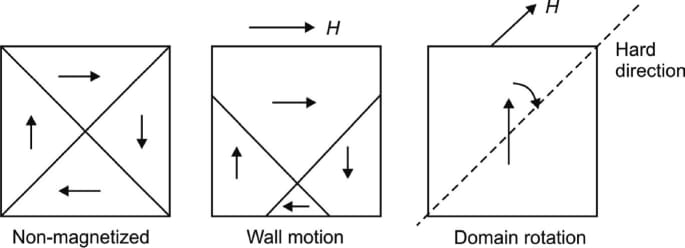}
    \caption{Domain structure}
    
\end{figure}
Single crystals' magnetization in a certain direction is always followed by a change in their physical dimensions. Depending on the crystal, it can stretch or contract along the magnetization direction. It contracts for Ni and expands for iron. We refer to this phenomena as magnetostriction.
Bloch wall thickness is limited, i.e., spin orientation progressively changes in the transition region (Fig.19). It is not endlessly tiny. Because the spin reversal is achieved in various steps, there is not much difference in spin orientation between two neighboring moments. As a result, the exchange energy associated to the wall decreases.
\begin{figure}[ht]
    \centering
    \includegraphics[width=1\linewidth]{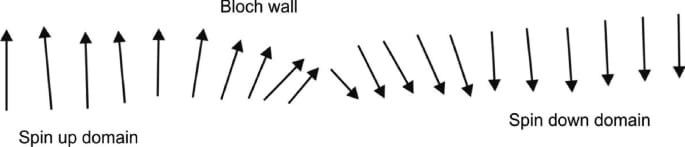}
    \caption{Bloch wall}
    
\end{figure}
It has been noted that in a ferromagnet, there are two crystallographic directions: one for easy magnetization,i.e.soft ferromagnetic materials, where saturation is obtained spontaneously, and another for hard magnetization,i.e.hard ferromagnetic materials, where saturation requires the application of an external field. The directions of easy and hard magnetization are [111] and [100] for nickel, respectively, whereas the opposite is true for iron. Magnetic anisotropy is the name given to this phenomenon. The magnetic anisotropy energy is the difference in energy between the easy and hard directions. Because more dipoles point in the hard direction when a wall is thicker, this energy has the effect of decreasing the thickness of the Bloch wall. Anisotropic energy favors a thin wall whereas exchange energy favors a thick wall. 
Little transverse domains are present at the end of the sample, as can be seen by closely examining the domain structure (Fig.20). These are known as closure domains because they further reduce magnetostatic energy by effectively completing the magnetic loop between two neighboring domains. "Bitter powder patterns," which are created by placing a drop of a colloidal solution of ferromagnetic particles on the specimen's meticulously prepared surface, serve as experimental proof that domains exist. Local magnetic fields close to the domain borders are quite strong. There, the particles are gathered, and a microscope may be used to view the domain.
\begin{figure}[ht]
    \centering
    \includegraphics[width=0.5\linewidth]{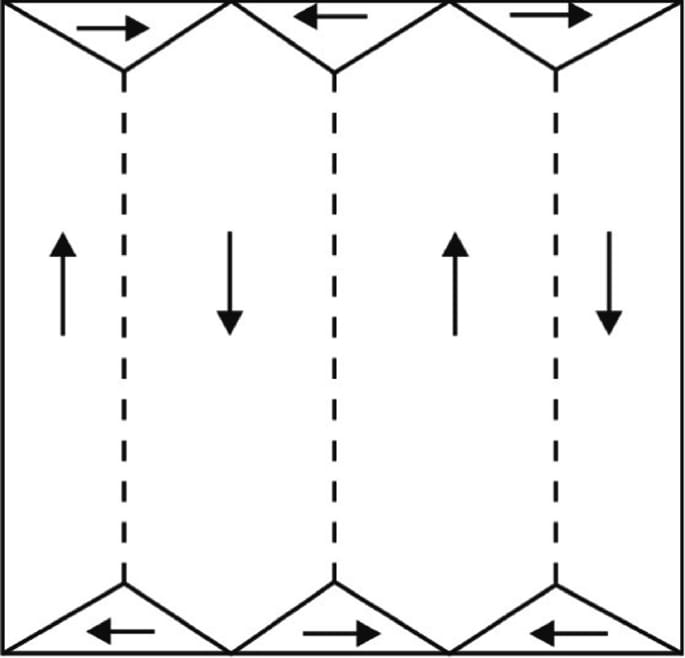}
    \caption{Closure domains}
    
\end{figure}
\subsection{Magnetism and hysteresis}
The spontaneous magnetization of ferromagnetic materials, like iron or magnetite, is the most amazing demonstration of magnetism in a solid. Typically, spontaneous magnetism is linked to hysteresis, a phenomena that James Ewing first observed and termed in 1881.For ferromagnets and ferrimagnets, flux density B and the field density H are not proportional. B changes as a function of H if the substance is originally unmagnetized.
The Curie point—a point at which a material's properties rapidly change—is used to describe ferromagnetic substances. The susceptibility follows a Curie-Weiss equation (Eq. ), roughly, and is independent of field strength above the Curie temperature. The behavior differs greatly below the Curie temperature. Very small fields can produce very large amounts of magnetization, and the relationship between magnetization and field strength is quite nonlinear. A typical plot of magnetic induction B as a function of field H in an iron sample is displayed in Figure 18.
\begin{figure}[ht]
    \centering
    \includegraphics[width=1\linewidth]{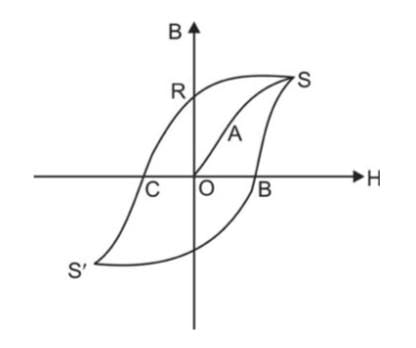}
    \caption{B versus H
hysteresis curve}
    \label{fig:enter-label}
\end{figure}
When a slowly rising field is given to initially unmagnetized iron, B travels along the OAS magnetization curve.The value of B is nearly constant at 1.5 Wb/m2 in a field of a few hundred ampere/meter. Reducing the magnetic field H causes the induction B to follow the path SR rather than the magnetization curve. B remains close to the saturation value even at H = 0, which corresponds to point R in Fig. 18. The retention of magnetization in zero field is referred to as remanence, and the value of B at this point is known as the residual induction.The value of B decreases and eventually becomes zero (point C) when a reverse field is applied; the field reverse induction value is then set up and rapidly approaches the saturation value. We refer to coercivity as the reverse field H at which B = 0. When a positive field is introduced and the reverse field is gradually eliminated, the induction eventually draws out the curve in the direction of S'BS. The hysteresis curve is the name given to this RCS'BSR curve. It demonstrates that variations in the applied magnetic field are always preceded by variations in the magnetic induction (B).Based on Weiss's domain theory, the B-H curve makes sense. The domains in an unmagnetized polycrystalline sample are randomly orientated, resulting in the absence of any magnetic moment in any direction. When a field is applied, domains with magnetization that is parallel to the field or at a slight angle to it grow at the expense of domains with antiparallel or nearly parallel magnetization, shifting the boundary between the two domains. At the beginning (OA in the B-H curve), the material's magnetization advances through minor (reversible) boundary displacements; however, the larger (irreversible) displacements are responsible for the steeper portion AB of the magnetization curve.Magnetization above the curve's knee occurs by rotation of the magnetization direction of overall domains; this is a challenging process with a comparatively sluggish rate of increase in magnetization. The magnetization stays relatively high until reverse fields are applied because there is limited change in the domain structure when the applied field reduces, which results in the hysteresis previously mentioned.The amountof work required to remagnetize a ferromagnet with a unit volume is proportional to the area of the B-H curve. During the remagnetization process, all of the work is converted to heat.

\bibliography{sn-bibliography}
\nocite{*}

\end{document}